\documentclass{elsart}

\usepackage{amsmath}

\begin{document}

\begin{frontmatter}

\title{It is possible for an observer to detect his motion through the space}
\author[Branislav]{Branislav Vlahovic\corauthref{cor}}
\ead{vlahovic@nccu.edu}
\corauth[cor]{Corresponding author.}

\address[Branislav]{North Carolina Central University, Durham, NC 27707, USA}

\begin{abstract}



One of two postulates that are base for special relativity is that the laws of physics are invariant in all inertial systems, which has as a consequence that it is impossible for an observer to detect his motion through space. It will be shown that this is in a contradiction with the results of the Hafele--Keating experiment, which established that time is going faster in an airplane going westward than in that going eastward, if compared with clocks located on Earth. The result of the experiment allows not only to conclude that Earth is rotating toward east, but also to calculate the speed of Earth motion. Performing similar experiments it is also possible to measure Earth speed around the Sun, its speed in our galaxy, and actually its absolute speed. To generalize this for any inertial frame and to explain why an absolute speed can be assigned to any inertial frame we introduced the triplets paradox.
\end{abstract}


\end{frontmatter}

\section{The triplets paradox}

	Let us consider identical triplets, one named A, who remains in a stationary coordinate system $K$ and two (B and C),  that make journey into space in high-speed rockets. Let make B and C go together with constant speed $V$ relative to the observer's A inertial coordinate system $K$.  Because of time dilation, since both brothers B and C are moving with their coordinate system $K'$ relative to the stationary observer A, observer A will find that both of his brothers B and C are aging more slowly, that their clocks (B and C) are having slower rate than his clock A. Lets now third brother decide to activate his rocket and move away from brother B with constant speed $V$ relative to the coordinate system $K'$. Again using the same argument, because of time dilation, since brother C is moving with his inertial coordinate system $K''$ relative to the observer's B coordinate system $K'$ with speed $V$, observer B will find that brother C is aging more slowly then he, that clock C of brother C is having slower rate than his clock B. It is important to underline that until now we did not specified any direction in which brothers B and C are moving relative to the brother A, and that we did not specified direction of the brother C relative to the brother B. Because all directions must be the same, there is no preferable direction in any inertial system \cite{AE}, so there is no reason to specify any direction. But is that statement correct, let see what will happen if we will specify directions of motion.

Let assume that brothers B and C are going along the $X$ direction in system $K$ and that latther brother C is going in the $X'$ direction in system $K'$, which is aligned with the $X$ direction of frame $K$. In this case all previous conclusions about the aging will be correct and brother A will see that brother B is aging at slower rate than he and that brother C will age at even more slower rate, that the clock C will have the smallest rate, smaller than  clocks A and B. 

But let now assume that brother C did not took direction $X$, that instead he is going in direction $-X$. Because, in this case brother C is actually stationary in the coordinate system $K$ his clock should have the same rate as the clock of brother A. This is already contradiction, because his rate should be smaller than rate of both brothers A and B (as it is obtained in previous case when he had direction $X$), since the difference in the clock rates between the brothers B and C should not depend on the direction which brother C took. Let us add here that since he is moving relative to the brother B that his clock rate should be smaller than the clock rate of clock B, which should be smaller than the clock rate of clock A. But his clock rate is the same as the clock rate of the clock A, since there is no relative speed between him and observer A. So, this establish contradiction. 

It appears that all directions in an inertial system are not the same. We have different results depending on which directions observer C took in inertial system $K'$. His clock will go faster than the clock of observer B if he is going in direction $-X$ and slower than the clock of observer B if he is going in direction $X$. But the system $K'$ is an inertial frame and all direction in that system should be the same, there should be no preferred direction. The time dilation should not depend about the direction which observer C will choose. There will be no such problem if observer B is stationary, if system $K'$ is not moving. So, one can use this method, sending rockets in all directions, to conclude if an inertial system is moving or not. If there is no difference in the rate of clocks, then the system is not moving, otherwise the system is moving in direction of the rocked that has the highest time dilation, the lowest clock rate. This is in contradiction with the first postulate of special relativity that prevents one to conclude if he/she is moving or not \cite{AE}. However, we will see that this triplets scenario is actually confirmed with the Hafele–-Keating experiment \cite{HK1},\cite{HK2}.

\section{Hafele–-Keating experiment and postulate on impossibility to detect motion of an inertial frame}

In 1971, Hafele and E. Keating took four cesium-beam atomic clocks twice around the world, first eastward, then westward, and compared the clocks against others that remained at ground. When reunited, the clocks were found to disagree with one another. They estimated special and relativistic time dilations performing simple calculations for the proper time integral: 
\begin{equation}
\tau_A - \tau_G = \int^t_0 [(\phi_A - \phi_G)/c^2 - (v_A^2-v_G^2)/2c^2]dt,
\end{equation}
where they used inertial system whose origin is attached to the center of freely falling Earth and which is non rotating. In this frame the vector velocity of aircraft is:
\begin{equation}
{\bf v}_A = {\bf v}_A^* + {\boldsymbol {\omega}} \times {\bf r}_A,
\end{equation}    
where t represent duration of the flight, $\phi_A$ and $\phi_G$ are Earth gravitational potentials at the aircraft and ground clocks positions,  ${\bf v}_A^*$ is the velocity of the aircraft with respect to the surface of Earth (positive for eastward motion and negative for westward), ${\boldsymbol {\omega}}$ is the angular velocity of Earth, and ${\bf r}_A$ is the radius vector to the aircraft. 
The clock traveling eastward was slowed down for $184 \pm 18$ ns due to the kinematics and clock traveling westward gained $96  \pm 10$ nanoseconds in comparison with the Earth clocks. These differences were consistent with the predictions of special and general relativity (due to the gravitational effects clock traveling eastward gained $144 \pm 14$ ns and clock traveling eastward gained $179  \pm 18$ ns).  The experiment has been reproduced by increased accuracy in 1975, 1976, \cite{Alley1}, \cite{Alley2}, 1996 \cite{NPL1}  and 2010 \cite{NPL2}. 
These experiments are confirmation of both general and special relativity predictions. However, as explained above they allowed us to detect Earth motion through space, which is in contradiction with the special relativity first postulate. One can even determine the speed of the inertial frame as in the above example speed of Earth surface, since values for all variables in the above equations can be determined. 

One can object that we did not determined the absolute motion of the Earth through space, but just the motion of the Earth surface. But in our example the Earth surface is our inertial system. Clearly it is not inertial frame, since
Earth is rotating around its axis and revolving around the Sun. However, this assumption does not have any impact on calculation of the kinematical part of time dilation in above equations or on conclusion related to the triplets paradox. 
In that inertial frame our clocks are in rest, and by the first postulate we should not be able to determine motion of our inertial frame.  However, using the same approach we can actually determine the motion of Earth around Sun or in our galaxy by sending rockets not around the Earth but in all directions radially away from Earth and measuring time dilations for each of them. The clock that will show the largest slow down, the smallest rate, will point to the direction of the Earth motion in space. The absolute velocity could be also estimated. For instance after one finds direction in which Earth is moving, then one can gradually increase speed of the rocket sent in opposite direction of the established motion until the maximum in the clock rate will be reached. The magnitude of that maximum time dilation will allow to calculate the Earth absolute speed $v_E$ using simple equation
\begin{equation}  
   {\tau_A-\tau_G} = {t_G}/{\sqrt{1-v_E^2/c^2}}.
\end{equation}
Assignment of an absolute speed to an inertial frame could have sense if one accept that any change of the speed changes property of the matter in the frame, for instance clock rate or rest mass. Basically, in this case system will be described by its history of acceleration. Consequently, if an absolute speed can be assigned to any inertial frame then obviously synchronization of clocks between any inertial frame also could be achieved, which also will impact definition of simultaneity between different inertial frames.   
\section{Conclusion}

Following procedure explained in the triplets paradox and Hafele-–Keating experiment one located in an inertial frame can determine if the frame is moving or not, by sending space rockets in all possible directions. If all rocket will show the same time dilation then that frame system is not only inertial, but it is in the absolute rest. If the inertial system is in any kind of uniform motion then one of the rockets, one which is in the direction of motion, will have the maximal time dilation, the slowest clock rate. From the information of the maximal clock rate achievable, one can also determine the speed of the inertial frame through space. This could be explained by the acceleration history of the inertial frame. If the system is in a motion it means that it went through an acceleration at one point of time. The speed acquired during that acceleration determines the property of the matter in that inertial frame, as for instance the time rate for clocks in that system or for instance the rest mass of the matter in that inertial frame.  This allows to define absolute speed or absolute rest frame, with frame zero speed, and also to synchronize clocks. Since each observers may agree about their absolute speed they also may agree about the time intervals or clock rates and sinhronize them accordingly.

  	  	\end{document}